\AtBeginDocument{%
  \providecommand\BibTeX{{%
    \normalfont B\kern-0.5em{\scshape i\kern-0.25em b}\kern-0.8em\TeX}}}

\documentclass[sigconf]{acmart}

\copyrightyear{2021}
\acmYear{2021}
\setcopyright{acmlicensed}\acmConference[ITiCSE 2021]{26th ACM Conference on Innovation and Technology in Computer Science Education V. 1}{June 26-July 1, 2021}{Virtual Event, Germany}
\acmBooktitle{26th ACM Conference on Innovation and Technology in Computer Science Education V. 1 (ITiCSE 2021), June 26-July 1, 2021, Virtual Event, Germany}
\acmPrice{15.00}
\acmDOI{10.1145/3430665.3456367}
\acmISBN{978-1-4503-8214-4/21/06}
\usepackage{multirow}
\usepackage{siunitx}
\usepackage{xspace}
\usepackage{paralist}
\usepackage{balance}
\usepackage{booktabs}
\usepackage{hyperref}

\newcommand{\snapcheck}{\textsc{SnapCheck}\xspace}%
\newcommand{\snap}{Snap\textsl{!}\xspace}
\newcommand{\scratch}{Scratch\xspace}%
\newcommand{\whisker}{\textsc{Whisker}\xspace}%
\newcommand{\bastet}{\textsc{Bastet}\xspace}%
\newcommand{\processing}{Processing\xspace}%

\begin{document}
\fancyhead{}

\title{\snapcheck: Automated Testing for \snap Programs}

\author{Wengran Wang}
\affiliation{%
  \institution{North Carolina State University}
  \city{Raleigh}
%   \state{NC}
  \country{USA}
  }

\author{Chenhao Zhang}
\affiliation{%
  \institution{Northwestern University}
     \city{Evanston}
%   \state{IL}
  \country{USA}
  }

\author{Andreas Stahlbauer}
\affiliation{%
  \institution{University of Passau}
     \city{Passau}
%   \state{Bavaria}
  \country{Germany}}

\author{Gordon Fraser}
\affiliation{%
  \institution{University of Passau}
        \city{Passau}
%   \state{Bavaria}
  \country{Germany}
  }

\author{Thomas Price}
\affiliation{%
  \institution{North Carolina State University}
  \city{Raleigh}
%   \state{NC}
  \country{USA}
  }

\renewcommand{\shortauthors}{}

\begin{abstract}
Programming environments such as \snap, \scratch, and \processing engage learners by allowing them to create programming artifacts such as apps and games, with visual and interactive output. Learning programming with such a media-focused context has been shown to increase retention and success rate. However, assessing these visual, interactive projects requires time and laborious manual effort, and it is therefore difficult to offer automated or real-time feedback to students as they work. In this paper, we introduce \snapcheck, a dynamic testing framework for \snap that enables instructors to author test cases with Condition-Action templates. The goal of \snapcheck is to allow instructors or researchers to author property-based test cases that can automatically assess students' interactive programs with high accuracy. Our evaluation of \snapcheck on~\num{162} code snapshots from a Pong game assignment in an introductory programming course shows that our automated testing framework achieves at least~\SI{98}{\percent} accuracy over all rubric items, showing potentials to use \snapcheck for auto-grading and providing formative feedback to students.

% Programming environments such as Snap, Scratch, and Processing engage learners by allowing them to create programming artifacts such as apps and games, with visual and interactive output. Learning programming with such a media-focused context has been shown to increase retention and success rate. However, assessing these visual, interactive projects requires time and laborious manual effort, and it is therefore difficult to offer automated or real-time feedback to students as they work. In this paper, we introduce SnapCheck, a dynamic testing framework for Snap that enables instructors to author test cases with Condition-Action templates. The goal of SnapCheck is to allow instructors or researchers to author property-based test cases that can automatically assess students' interactive programs with high accuracy. Our evaluation of SnapCheck on 162 code snapshots from a Pong game assignment in an introductory programming course shows that our automated testing framework achieves at least 98% accuracy over all rubric items, showing potentials to use SnapCheck for auto-grading and providing formative feedback to students.
\end{abstract}

\maketitle

\section{Introduction}

Visual, interactive programming projects, such as creating student-designed apps and games, are widely used in many introductory programming courses (e.g., \cite{garcia2015beauty, guzdial2003computer}). They encourage students to pursue projects that produce a computational artifact that they can interact with, and to express their ideas creatively, which has been shown to motivate students \cite{guzdial2003computer}. Additionally, many popular block-based programming environments, such as \scratch \cite{maloney2010scratch} and \snap \cite{harvery2012byob}, are specifically designed to create such visual, interactive programs.

One challenge to using visual, interactive programs in the classroom is that the very properties that make them engaging also make them difficult to assess automatically. While traditional introductory programming tasks can usually be tested using simple input/output pairs, visual, interactive programs are controlled by sequences of user inputs (e.g., key presses and mouse clicks), and their output is based on the visual composition of sprites on a screen\footnote{A sprite in \snap~ is an object (such as in object-oriented programming) and can have its own code (scripts), and variables.}, making it difficult to write automated test cases. Not only does this make it time consuming for instructors to assess these programs, limiting the ability to scale up courses, it also precludes the use of the automated, formative feedback found in many introductory courses. Prior work addressed these problems by developing automated, functional tests for \scratch \cite{stahlbauer2019testing}. However, it has not been extended to automated assessment for \snap programs.

In this work, we introduce \snapcheck, an automated testing framework for visual, interactive programs. We define how to author \snapcheck test cases with a domain-specific-language (DSL), and describe how an instructor can use \snapcheck's user interface to author test cases with  Condition-Action templates that simulate the interaction rules of a human tester when running a \snap program. We evaluated \snapcheck on 162 program snapshots and final submissions from 42 students working on a programming assignment. Our results suggest that \snapcheck can accurately test these programs by providing at least ~\SI{98}{\percent} accuracy on all rubric items, showing that \snapcheck can be used by instructors to automatically assess students' visual, interactive programs in \snap.

\newcommand{\boldparagraph}[1]{\noindent \textbf{#1}}%

\section{Related Work}

\boldparagraph{Automated Assessment.}
Effective feedback should be timely and specific \cite{shute2008formative, Scheeler2004}. However, in today's growing-size CS classrooms~\cite{computing2017generation}, traditional, manual assessment is usually insufficient for offering timely feedback to all students; many K-12 in-service programming instructors are in the process of transitioning from non-programming backgrounds, and therefore lack proficiency in grading students' assignments manually \cite{ericson2014preparing}.

Automated assessment not only reduces instructors' grading efforts, but also allows students to receive automated feedback on where they were correct or made mistakes on \cite{edwards2017codeworkout} in the middle of programming. Unlike instructor feedback, automated feedback can be easily propagated to all students having the same mistake, allowing a larger group of students to benefit from receiving feedback. This feedback can take the form of success or failure of test cases \cite{hovemeyer2013cloudcoder, edwards2017codeworkout, wang2020crescendo}, highlighting erroneous code \cite{price2017edm, edmison2017}, or identifying likely misconceptions \cite{gusukuma2018}. Providing such immediate feedback has been shown to engage and motivate students \cite{marwan2020icer}, improve their learning outcomes~\cite{corbett2001locus}, and does not pose social threat to students \cite{price2017icer}.

To build automated assessment tools, researchers used two prevalent methods to analyze students' programs: syntax-based static analysis (e.g., \cite{wang2020crescendo, marwan2020icer, price2017edm, piech2015learning}), and functional program analysis, (e.g., \cite{patil2010automatic,hovemeyer2013cloudcoder, kumar2006explanation, edwards2017codeworkout}), generated from dynamically running the program rather than inspecting the structure of its code. 
%We discuss these two methods below. 

\boldparagraph{Syntax-Based Program Analysis.}
In block-based languages, many students complete programming assignments by making apps, games and simulations \cite{garcia2015beauty}. Because of the visual and interactive output of these programs, it is challenging to perform functional tests that checks correctness of procedures based on input-output pairs in these programs (we discuss \textit{why} in Section~\ref{sec:formative_study}). Therefore, many programming analysis approaches made use of syntax-based program analysis to check students' program structures based on abstract syntax tree (AST) rules \cite{wang2020crescendo, marwan2020icer, ball2018lambda}. 

For example, \citeauthor{marwan2020icer} developed a data-driven syntax-based autograder in \snap, to provide real-time adaptive feedback, and found that it increased students' engagement with the programming environment and intentions to persist in CS \cite{marwan2020icer}. Crescendo is a \snap-based self-paced learning tool used syntax-based rules to automatically grade rubric-like assignment requirements (e.g., draw a triangle) during programming. This rule-based syntax-based analysis checks AST substructures to determine whether the code satisfied specific programming requirements (e.g., is a ``move'' block inside of a ``do repeat'' loop?).
In the evaluation, they found that students benefited from receiving immediate feedback, as shown by high completion rates and fast task completion. However, such syntax-based rules can be brittle and difficult to author, and may lead to assessment errors when not aligning with a student's solution strategy \cite{wang2020crescendo}. This suggests that checking the syntactic structure is insufficient to correctly assess student projects.

\boldparagraph{Functional Program Analysis.} 
In text-based programming languages such as Java and Python, instructors and learning systems commonly use functional testing to automatically assess students' programming assignments. For example, CloudCoder \cite{hovemeyer2013cloudcoder}, Problets~\cite{kumar2006explanation} and CodeWorkout \cite{edwards2017codeworkout} included test-case-based program assessment in its built-in programming assignments, enabling students to receive immediate feedback upon completion of a programming assignment. Marmoset \cite{spacco2006experiences} allows students and instructors to author and use test cases to grade their assignments, and provides instructors with overviews of students' performance on instructors' tests. These systems make use of \textbf{functional tests}, where certain test cases that include input-output pairs were executed, to examine the correctness of a certain procedure (i.e., usually a function with a fixed name) inside student programs. Prior work used functional tests to provide automated feedback to students, which has been shown to improve their learning outcomes \cite{corbett2001locus, gusukuma2018misconception, mitrovic2003intelligent}. For example, \citeauthor{gusukuma2018misconception} used functional analysis to find students' erroneous programs in Python, and generate misconception-driven feedback to those programs, which has improved students' performance. However, these traditional, input/output-based testing may not apply to visual, interactive programs. 

\whisker~\cite{stahlbauer2019testing} is an automated testing framework for visual, interactive programs written in \scratch. Similar to \snap programs, \scratch programs are highly concurrent, rely heavily on timers, and are driven by events such as user interactions using keyboard and mouse inputs~\cite{stahlbauer2019testing}. 
To address the challenges of visual, interactive programs, \whisker allows educators to write and to automatically generate~\cite{deiner2020search} test cases in JavaScript which simulate user inputs to a \scratch program and  observe the program's resulting behavior. The authors evaluated \whisker on 37 students' programming assignments on their completion of 28 properties, such as ``only one apple must fall down at a time.'' They found the results produced by \whisker to be strongly correlated with the instructors' grading, showing that it is possible to use functional tests to grade student projects in \scratch, although the actual test-authoring may still be challenging for instructors~\cite{stahlbauer2019testing}.
\snapcheck is inspired by \whisker, but targets \snap.
More elaborated steps towards checking \scratch programs are taken by \bastet~\cite{Bastet},
which, for example, implements an exhaustive state-space exploration to check for 
violations of requirements.

\section{Formative study \& Design Challenges}
\label{sec:formative_study}%

Our goal is to allow programming instructors to automatically assess \snap programs. \snap allows students to create complex games and apps using visual and block-based programming; it is used by many students in the AP CS Principles and university course each year as part of the BJC curriculum \cite{garcia2015beauty}. To understand the design opportunities and challenges when grading students' visual, interactive assignments, we started by manually inspecting \num{42}~student submissions, taken from a classroom assignment. In this assignment, students were required to implement a one-player Pong game, where there is one paddle on the left / right side of the stage \footnote{A stage is where \snap displays its sprites and actions}, which must bounce a ball against a wall. The instructor required students to implement the game with the following 10 requirements:
\begin{enumerate}
\item \emph{key\_up}: The paddle moves up with an up arrow key.
\item \emph{key\_down}: The paddle moves down with a down arrow key.
\item \emph{upper\_bound}: When touching the upper bound, the paddle does not move upwards  when the up arrow key is pressed.
\item \emph{lower\_bound}: When touching the lower bound, the paddle does not move downwards even when the lower key is pressed.
\item \emph{space\_start}: The ball starts movement when space key is pressed. 
\item \emph{edge\_bounce}: The ball moves and bounce on edge unless touching the back wall.
\item \emph{paddle\_bounce}: The ball bounces back to stage when touching paddle.
\item \emph{paddle\_score}: If the ball touches the paddle, increase score.
\item \emph{reset\_score}: If the ball touches the back wall behind the paddle, reset score to 0.
\item \emph{reset\_ball}: If the ball touches the back wall behind the paddle, reset ball to the center of the stage.  
\end{enumerate}

We manually analyzed these \num{42}~student programs, and identified 5 design challenges of creating automated testing framework for these visual, interactive programs, described below.

\boldparagraph{Dynamic User Inputs.}
For non-interactive programming problems, students' programs usually consist of one or more individual functions.
A standard testing approach is to use test cases that specify input-output pairs, checking if the output of the function matches the expected output for each input. 
For example, in an integer sorting problem, the students write a function (such as \texttt{sort()}) that takes a list of integers as the input parameter. 
To test students' programs, the instructor prepares test cases which call the students' \texttt{sort()} functions with a pre-defined list of integers, and then check whether the values returned are identical to the corresponding sorted list. 

For visual, interactive programs like Pong, however, the input to the program is a constant stream of signals from input devices like the keyboard and mouse. Users observe the changes of graphical elements and send different inputs according to what they observe over time. Such inputs are dynamic and dependent on the program state. For example, in the Pong program, a user may press the up arrow to move the paddle up when they see the ball go up. Such input stream can be challenging to encode as standard input data.

\boldparagraph{Visual Outputs.}
In a non-interactive programming problem (e.g., an integer sorting problem), the instructor may define the outputs as a list of integers and check whether the student program returns such output. However, it is less clear how to define output as a single, ``correct'' value for visual, interactive programs. These programs include multiple elements (e.g. the sprites in \snap or Scratch), each having its own properties (e.g., direction, position).

\boldparagraph{Delayed Responses or Outputs.}
For visual, interactive programs, specifying just the input and output is  insufficient---the output sometimes happen after certain delays, and this time difference may be different across student programs. For example, when testing the \textit{key\_up} behavior, with different implementation approaches, the paddle movement and the keyboard pressing happen at different timestamps: a student's program may move the paddle with key continuously: the paddle may move smoothly upward, or in larger bursts. The delays between inputs and outputs are different in these two scenarios, and caused difficulties to specify the delay in a test program. 

\boldparagraph{Requirements with Temporal Constraints.}
When specifying test cases for non-interactive programs, instructors do not need to distinguish between which specifications are ``forever true'', and which are ``sometimes true''. For example, in the integer sorting program, the program finishes execution in milliseconds, and instructors do not need to specify intermediate invariants during the program execution, such as the order of the array after the third iteration of the algorithm.

However, requirements on interactive programs include different temporal constraints: some requirements should be \textit{always} satisfied (e.g., \textit{edge\_bounce}); some should be checked only once (e.g., \textit{space\_start}); some should be checked after another (e.g., \textit{upper\_bound} should be checked after \textit{key\_up})). Specifying such time constraints adds challenges.

\boldparagraph{Various Implementations.}
We have identified several challenges of doing functional testing in these interactive programs, such as challenges to specify input, output, and specify temporal restrictions. Prior work in automatic assessment of \snap programs identified similar challenges, and instead applied syntax-based rules on the AST \cite{wang2020crescendo, marwan2020icer, ball2018lambda}. However, our analysis of students' Pong code shows that students included a variety of code patterns to implement the same behavior, causing challenges to apply rule-based approaches directly to check for AST subcomponents: 
%https://app.diagrams.net/#G1D2QoGBMxrraBDzfDGvv_TyNDAQgyZf7W 
\begin{figure}
   \centering
   \includegraphics[width=.35\textwidth]{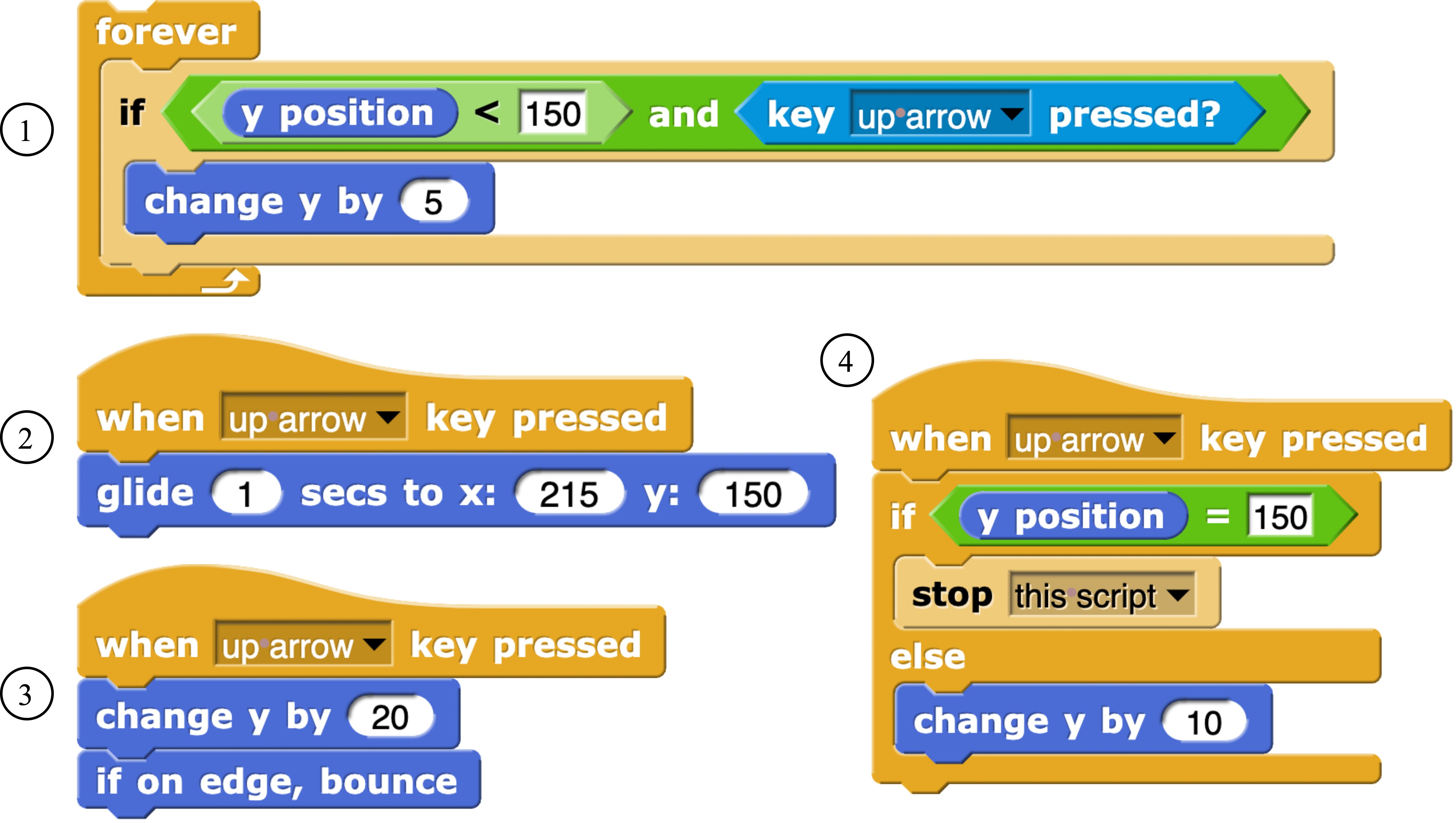}
   \caption{Examples of ways students constructed code to allow paddle to not move with the upper arrow key when it hits the upper bound.}
   \label{fig:rowOfHills}
   \vspace{-1em}
\end{figure}
For example, Figure \ref{fig:rowOfHills} shows that when implementing the \textit{upper\_bound}, students may 
1) stop sprite movement when its $y$~position is larger than a certain value; 
2) use \snap's built-in ``glide'' block to fix the target of sprite movement; 
3) use \snap's built-in block ``if on edge, bounce'' to stop sprite movement when it hits edge; 
4) stop the entire script started by a ``when up arrow key pressed'' hat block. 
Students may also include combinations of the above approaches, resulting in multiple unpredictable ways to construct code that implements the same observable behavior. This caused syntax-based analysis approaches to be insufficient to understand and analyze these programs. 

These design challenges are in line with the challenges addressed by \whisker~\cite{stahlbauer2019testing} in the context of testing \scratch programs. We next discuss the design of \snapcheck to address these goals. 

\begin{figure*}
  \includegraphics[width=.9\textwidth]{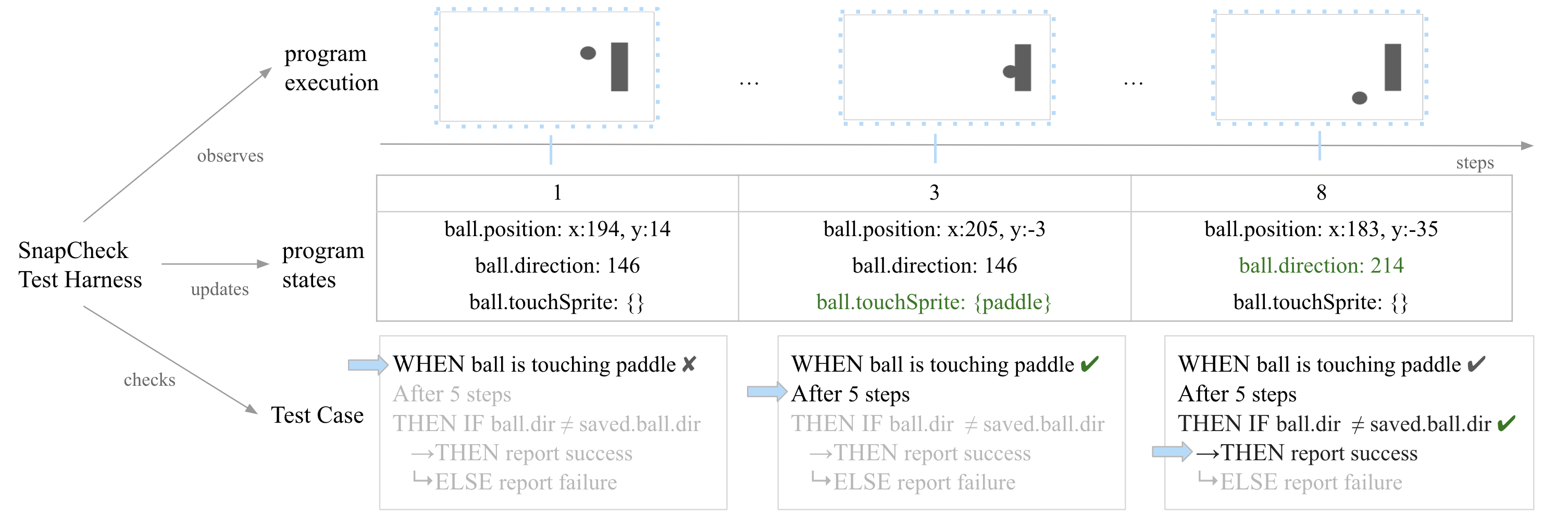}
  \vspace{-0.5em}
  \caption{Automated Testing for \textit{paddle\_bounce}: the \snapcheck Test Harness keeps updating program states. When the \textit{WHEN} condition evaluates to True (Step 3), it checks for the \textit{THEN-IF} condition after the defined delay (Step 8).}
   \label{fig:stepping}
  \vspace{-0.5em}
\end{figure*}
\section{Automated Testing with \snapcheck}

Based on the challenges uncovered by the formative analysis, we formulate the design goal for \snapcheck as to provide automated testing framework for \snap, with user-friendly specification of desired properties, including complex inputs, outputs, and temporal restrictions. We implemented \snapcheck in \snap, but the testing model of \snapcheck can also be applied to interactive sprite-based visual programs on other systems, such as \scratch \cite{resnick2009scratch}. This would require representing the relevant objects (e.g., Sprites) and properties (e.g., x coordinate) of those systems. A similar model may be able to support 3D interactive graphical programs such as Alice~\cite{cooper2000alice}. \snapcheck is an open-source software that can be downloaded at \href{https://github.com/emmableu/SnapCheck}{\color{cyan}{github.com/emmableu/SnapCheck}}.

\subsection{Authoring Test Cases in \snapcheck}
\label{sec:specify}
\begin{figure}
   \centering
   \includegraphics[width=.43\textwidth]{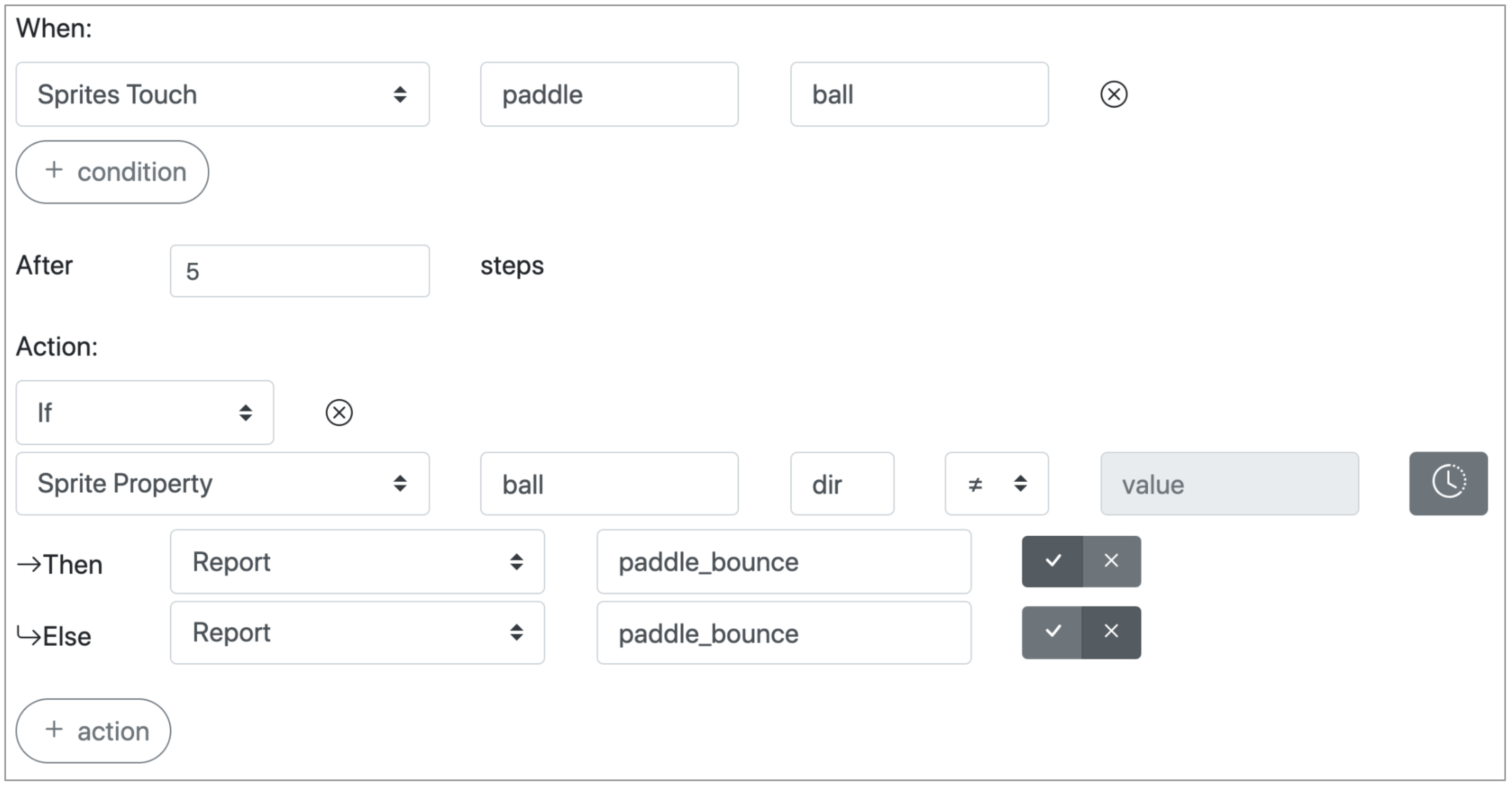}
   \caption{Using \snapcheck UI to specify \emph{paddle\_bounce}}
   \label{fig:ui}
%   \vspace{-2em}
\end{figure}

To test programs using \snapcheck, an instructor needs to write a list of \textit{test cases}, each with a name, and its rules to programmatically imitate what a human tester does when running a \snap program, described below. The test cases can be defined either through an selection-based User interface (shown in Figure \ref{fig:ui}), or in JavaScript by using APIs provided by \snapcheck. For example, to check \textit{paddle\_bounce} rubric item, an instructor may define a test case using the same name --- shown in Figure \ref{fig:ui}.

    \begin{quote}
      (\textit{always}:);
       \textit{WHEN} ball is touching paddle;
       \textit{After} 5 steps;
       \textit{THEN IF} the direction of paddle is different from its saved direction;
       \textit{THEN} report test success;
       \textit{ELSE} report test failure.
    \end{quote}
    
\begin{figure}
\includegraphics[width=.48\textwidth]{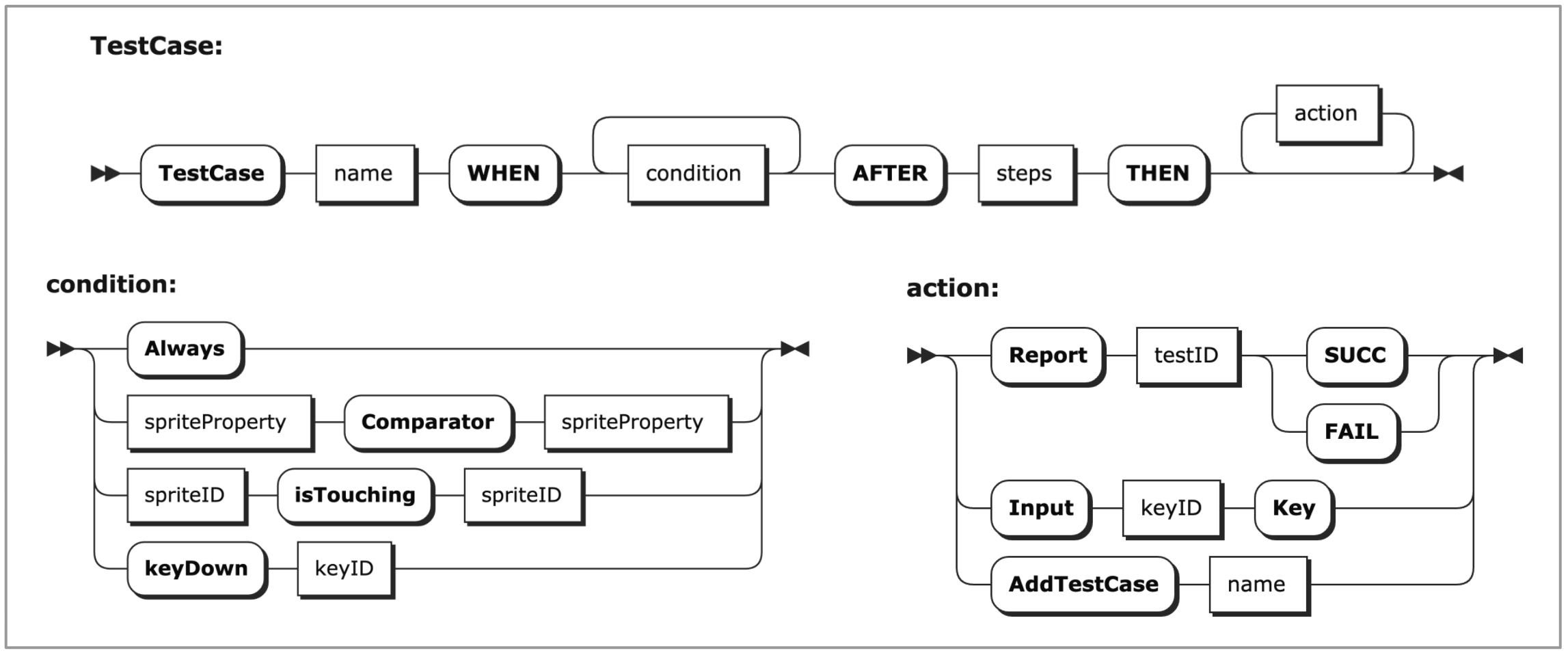}
\vspace{-2em}
\caption{Grammar of the \snapcheck DSL}
\label{fig:grammar}
\vspace{-1em}
\end{figure}

We next define how such requirements can be specified step by step, and illustrate the \snapcheck grammar using a railroad syntax diagram, shown in Figure \ref{fig:grammar}.

\vspace{0.5em}
\boldparagraph{Step 1: Define a \textit{WHEN} condition: } \textit{WHEN Ball touches paddle.}

The first step is to define \textit{when} the rubric item should be evaluated, which is done using a WHEN condition.

A \textit{condition} is a predicate function \footnote{A predicate function is a function that returns True or False.} that keeps track of the program state (e.g., sprite location, variable values) during the program execution. It evaluates to true if a program state satisfies a required property. \snapcheck allows multiple options for specifying a condition, including 1) sprite location and directions; 2) relation between sprites, such as one sprite touches another or a sprite touching the edge of the stage; 3) variable values; 4) comparison with a cached state saved from from the last step of execution, such as x coordinate smaller than that from 1 step ago, meaning the sprite moves to the right. 

A \textit{WHEN condition} is a predicate function that triggers the start of a test case when it is true. For example, if an instructor defines a \textit{WHEN condition}: \textit{WHEN ball touches paddle}, \snapcheck will constantly track whether the ball is touching paddle, and when this function returns to true, it starts executing the next statement defined by the user, described in Step~2.

\textit{Implementation detail: } How does \snapcheck define and track its \textit{WHEN conditions} internally? As shown in Figure~\ref{fig:stepping}, \snapcheck uses a \textit{Test Harness} module to monitor the program state at every internal execution step of \snap (i.e., a block of code). At every step, the \textit{Test Harness} executes the \textit{WHEN condition} function, and when it returns true, it starts executing the test cases following the \textit{WHEN condition}, which we describe next.

\vspace{0.5em}

\boldparagraph{Step 2: Define a \textit{THEN-IF} condition: } \textit{THEN IF ball changes direction.}

After defining the \textit{WHEN condition}, the instructor can use the keyword \textit{THEN-IF} to specify correct (or incorrect) behavior of the program when a rubric item is evaluated, as determined by the WHEN condition. This property is defined using the same set of predicate functions as defined in Step~1. 

Here, if the instructor wants to check whether the ball changes direction, they can define the THEN-IF condition to check whether the ball's direction has changed from its ``saved direction'' - In \textit{THEN-IF} conditions, all properties have a \textit{saved} variant, which is by default the value of that property (e.g., direction) saved from Step~1 in the \textit{WHEN} condition. Based on whether the \textit{THEN-IF} condition is satisfied, we may next run Test Case executes actions, defined in Step~4.

\textit{Implementation detail: } The keyword \textit{THEN} starts a callback function, which begins executing after the previous condition is satisfied. It may optionally take the saved ball direction as an argument. When called, the callback function compares the current ball direction with the argument, and records the success and failure respectively. In addition, because \textit{THEN} starts a callback function, it is easily to add multiple \textit{THEN} conditions after one, to afford more expressive test authorizations.

\vspace{0.5em}
\boldparagraph{Step 3: Define a delay:} \textit{After 5 steps.}

The \textit{WHEN condition} and the \textit{THEN-IF condition} do not always happen simultaneously: there can be delays between two conditions, explained in Section~\ref{sec:formative_study}. This delay happens because it takes time for the student program to execute from one line (e.g., detection of touching) to another (e.g., change the ball direction), especially when they are not directly adjacent. 

An instructor may define the delay between the \textit{WHEN condition} and the \textit{THEN-IF condition} based on how many steps will be executed between these two conditions. Here, a ``step'' is a time interval, usually in milliseconds. Program environments such as \snap and \scratch make use of a ``step'' to update sprite properties atomically on the stage \cite{stahlbauer2019testing}, which by default takes place at every frame of the display. For example, Figure \ref{fig:stepping}, \snapcheck executes ``After 5 steps'' after the \textit{WHEN condition}, meaning to check the \textit{THEN-IF condition} after 5 steps of program execution. 

\textit{Implementation detail: } \snapcheck uses a countdown to track how many steps has passed between the \textit{WHEN condition} and the \textit{THEN-IF condition}:  After the \textit{WHEN condition} evaluates to true, the paddle direction referred to in the callback function is saved at that step. The test case then waits for the delay parameter number of steps and executes the callback defined in Step~2.

\vspace{0.5em}
\boldparagraph{Step 4: Define a \textit{THEN} action:} \textit{THEN report test success.}

An \textit{action} defines programmatic actions \snapcheck can do to 1) report the success or failures of a test case, 2) supply input to the executing program, or 3) change how future tests are run. For example, an action can be reporting a test success; it can also be to used to remove or add another test case test case, which we describe in detail in Step~5. 
 
After executing ``\textit{WHEN} ...,  \textit{After} ..., \textit{THEN-IF} ...'', The instructor can use the keyword \textit{THEN} to define what to report, such as test success or failures. This creates a data point for the final test statistics, which writes to a spreadsheet that gets generated at the end of the tests. 

\textit{Implementation detail: } As in Step~2, the keyword \textit{THEN} also starts a callback function that follows the previous callback functions (e.g., in Step~2), or a direct \textit{WHEN condition} (e.g., in Step~1). Here, the callback reports the test to be a success. 

\vspace{0.5em}
\boldparagraph{Step 5: Specify temporal constraints:} \textit{Always.}

As explained in Section~\ref{sec:formative_study}, some rubric items are satisfied when a student's program satisfied the requirements once, while some require that it satisfies the requirement the whole time. To allow specification for different time constraints, an instructor may use 3 ways to specify temporal constraints:
(1)~ \textit{Always.} v.s. \textit{One-shot.}: The default is \textit{Always.}, where \snapcheck runs the test case non-stop, even when it returns a success;  \textit{One-shot} test cases, on the other hand, only check the test case once. (2)~\textit{add-on-start}: the instructor may use the keyword \textit{add-on-start} to define test cases that test programs immediately after the program starts.
(3)~\textit{Add/remove test cases}: As explained in  Section~\ref{sec:formative_study}, some test cases have dependencies, and may benefit from specifying test cases with orders, such as testing \textit{space\_start} first, and \textit{paddle\_bounce} next. Therefore, in the action part of the program (e.g., in Step~4), an instructor may also define adding and removing test cases, to specify the sequential order of test cases.

\subsection{Defining Test Inputs}
\label{sec:follow_ball}
When testing visual, interactive programs, instructors may need to send relevant inputs reacting to what is observed. These inputs can also be encoded as test cases using the \emph{THEN input...} syntax, where ``Input'' is a type of \textit{THEN action} defined in Step~4. For example, in the Pong game, a user needs to press the up/down arrow keys when they observe that the ball is in a higher/lower position than the paddle in order to make the paddle follow the ball, so that touching can be observed. We call this ``follow ball' input, which can be directly encoded as a test case in \snapcheck:
\vspace{-0.1em}
    \begin{quote}
        \textit{WHEN} the y coordinate of the ball is greater than that of the paddle;
        \textit{After} 1 step;
        \textit{THEN} input `up arrow' key for 4 steps.
    \end{quote}
\vspace{-0.5em}

\section{Experiment}

\boldparagraph{Dataset \& Experiment Setting.} To evaluate how well \snapcheck achieves its goal of offering accurate automated tests for students' visual, interactive programs, we used the same dataset as in our formative study. But in addition to the 42 students' final submissions in Pong,  we also made use of their their intermediate snapshots, based on their trace data at 10 minutes, 20 minutes, and 30 minutes (42, 40, and 38 snapshots at each time, respectively). We manually graded all 162 programs based on the same 10 project requirements specified in Section~\ref{sec:formative_study}. 
%
%, but also included students' intermediate project trace data, where we collected at every timestamp when the students make a change to their code. 
%
We selected Pong as a proper assignment to test the affordances of \snapcheck, since it includes a wide range of properties, can be implemented using a variety of different approaches, and its requirements for test automation embody the challenges that we designed \snapcheck to address.  
One researcher authored three different input series for \snapcheck: 1) up arrow key; 2) down arrow key; 3) ``follow ball'' (see Section~\ref{sec:follow_ball}, and when ball touches paddle, stops following.
We authored automated test sequences to re-execute the program by clicking green flag to start game, and clicking space key to start ball movement. Because many students starts ball movement by pointing to a random direction, with each input sequence, we specify three different random seeds, and run program against all test cases using all combinations of random seeds and input series. 
At the end of testing, \snapcheck reports the \#satisfied reports / \#total reports (i.e., satisfaction rate) during each run. Because some test cases may have spurious failures ( e.g., when \textit{upper\_bound} is satisfied, \textit{key\_up} is not satisfied), we used satisfaction rates that were lower than 0.1 to indicate a failure of a test case, and those higher than 0.1 being satisfied test cases \footnote{We set this threshold a priori, but our final analysis found that a threshold of 0.05 also reached similar results, showing that \snapcheck catches the satisfaction of properties most of the times.}.

\boldparagraph{Test Accuracy.} Table~\ref{table:accuracy} shows the accuracy, precision, and recall of \snapcheck's grading on 162 students' snapshots. Our results show that \snapcheck was able to provide accurate grading on all rubric items. The high accuracy of these graded assignments showed that teachers may benefit from \snapcheck for  automatically grading students' project submissions. While instructors might ideally want 100\% accuracy, we argue that most teaching assistants (TAs) also do not grade students' work with perfect accuracy, and instructors rely on students to dispute incorrect grades due to oversights.

Our results on students' intermediate, incomplete programs suggest there may also be applications of this work to providing formative feedback as students are working. While auto-grading requires a set of test inputs and time to run them, a student may potentially use \snapcheck to run their program and view the test results presented by \snapcheck, such as ``complete'', ``incomplete'' or ``not demonstrated by the test run''. This may help students develop productive testing strategies, as prior work suggests that they often do not use systematic testing \cite{dong2019defining}.

\begin{table}[]
\caption{Accuracy, precision, and recall of \snapcheck’s
functional tests in 162 intermediate and final programming
snapshots, on 10 rubric items. As well as the prevalence of
positive items within each rubric item.}

\begin{tabular}{@{}rrrrrr@{}}
\toprule
rubric item    & prevalence & accuracy & precision & recall & F1   \\ \midrule
key\_up        & 0.76       & 0.99     & 0.99      & 1.00   & 1.00 \\
key\_down      & 0.73       & 0.99     & 0.98      & 1.00   & 0.99 \\
upper\_bound   & 0.55       & 1.00     & 1.00      & 1.00   & 1.00 \\
lower\_bound   & 0.56       & 0.99     & 0.99      & 1.00   & 0.99 \\
space\_start   & 0.50       & 0.98     & 0.98      & 0.99   & 0.98 \\
edge\_bounce   & 0.49       & 1.00     & 1.00      & 1.00   & 1.00 \\
paddle\_bounce & 0.42       & 0.99     & 0.97      & 1.00   & 0.99 \\
paddle\_score  & 0.35       & 0.99     & 1.00      & 0.98   & 0.99 \\
reset\_score   & 0.23       & 1.00     & 1.00      & 1.00   & 1.00 \\
reset\_ball    & 0.31       & 0.99     & 0.98      & 1.00   & 0.99 \\ \bottomrule
\end{tabular}

\label{table:accuracy}
\end{table}

\section{Discussion}
Our results show that \snapcheck can not only assess students' final projects accurately, but also their incomplete programs, which includes buggy and erroneous code. This suggests the potential of using \snapcheck to offer automated feedback to students on their progress \textit{during} programming. Similar to prior work~\cite{stahlbauer2019testing}, our work identified challenges in automatically assessing students' visual, interactive programs. Additionally, we present a novel testing framework for \snap. Going beyond prior work such as \whisker, we designed a selection-based user interface and presented a step-by-step procedure to author test cases. 

In addition, our experience with \snapcheck also identified \textbf{key factors that influence the accuracy of performing automated tests for visual, interactive programs}. These are factors that any functional tests on different types of visual, interactive programs (e.g., Scratch) would also potentially rely heavily on, and therefore add important considerations for future work, discussed below:

\boldparagraph{High-Coverage Inputs.}
\snapcheck is only able to check for the presence of behaviors when they are \textit{executed}. For example, a \textit{paddle\_bounce} behavior would not be present if the paddle does not catch the ball, potentially leading \snapcheck to make false negative predictions if this behavior is implemented. In this assignment, we designed the \textit{follow ball} input to allow the paddle to follow the ball's movement and catch it, which requires expert knowledge of the program. However, in more open-ended assignments, one set of input series may not cover all possible student programs, and may cause low program coverage and lower testing accuracies. 

\boldparagraph{Temporal Specifications.}
The \snapcheck DSL uses a \textit{delay} to specify the time difference between the \textit{WHEN} condition and the \textit{THEN-IF} condition. A delay can take any number of steps of the program execution, defined by the user. Because each step executes in milliseconds, the difference between 1 step and 10 steps may be undetectable to the human eye, but it can make an important difference in the detection of behaviors. In our experiment, we generally used a higher number of steps in delays (e.g., for the \textit{paddle\_bounce} rubric item,  even if the ball changes direction immediately when touching paddle, we detect the change of direction \textit{after} 5 steps). However, this approach may not work if another event changes the ball's state in before 5 steps.

\boldparagraph{Complex Conditions.}
Unlike describing a rubric item in natural language, testing programs requires the author to specify details formally, including specifying multiple, complex conditions to handle different edge cases. In the Pong program, there are two examples: 

1) The rubric item \textit{space\_start} says that the ball should not move before the space key is pressed (i.e., the ball movement is triggered by the user pressing the space key, not immediately when the game starts). Therefore, an instructor needs to define test cases to first check that the ball does not move when the game starts, and after key presses, start checking for ball movement. In addition to the complex conditions, this test case may still fail erroneously, e.g., when a student's program first resets from another position to the center when the game starts, the ``not moving when game starts'' check would fail, causing false negatives in the detection.

2) For the rubric item \textit{paddle\_bounce}, a student may fail to implement this behavior, so the ball passes the paddle but bounces off the wall behind the paddle. Distinguishing these two different bouncing behaviors requires a specification of the difference in delays, the x coordinates of the ball in relation to the paddle, and the change of ball y coordinates in between \textit{WHEN} condition and \textit{THEN} condition. Authoring these specifications require knowledge of the Pong program, and may easily cause overly strict / loose specifications. Going beyond Pong, such requirements of complex conditions may cause difficulties to specify more complex behaviors, such as shooting or jumping, which require authoring chains of multiple test cases.

\section{Conclusion}

We introduced \snapcheck, a novel, automated testing framework that tests visual, interactive programs in \snap. We explained how to author test cases in \snapcheck using a domain-specific-language, and presented a novel UI for authoring these test cases. Our evaluations on 162 student projects show that instructors may use \snapcheck to accurately assess students' programs, and to enable offering formative feedback in the middle of programming.

\section{Acknowledgements}
This material is based upon work supported by the National Science Foundation under Grant No. 1917885.

\balance
\bibliographystyle{ACM-Reference-Format}
\bibliography{sample-base}


\end{document}